\documentclass[a4paper,12pt]{article}
\usepackage{epsfig}
\begin{document}

\begin{center}

{\large\bf Parameter-independent predictions for nuclear shapes

 and B(E2) transition rates in the proxy-SU(3) model} \bigskip
 
 \footnotesize

A. Martinou$^1$, S. Peroulis$^1$, D. Bonatsos$^1$, I. E. Assimakis$^1$, S. Sarantopoulou$^1$, 

N. Minkov$^2$, R. B. Cakirli$^3$, R. F. Casten$^{4,5}$, and K. Blaum$^6$

\medskip

{\it 
$^1$ Institute of Nuclear and Particle Physics, National Center for Scientific Research  ``Demokritos'', Athens, Greece 

$^2$ Institute of Nuclear Research and Nuclear Energy, Bulgarian Academy of Sciences, 

72 Tzarigrad Road, 1784 Sofia, Bulgaria

$^3$ Department of Physics, University of Istanbul, Istanbul, Turkey 

$^4$ Wright Laboratory, Yale University, New Haven, Connecticut 06520, USA  

$^5$ Facility for Rare Isotope Beams, 640 South Shaw Lane, Michigan State University, East Lansing, 

MI 48824 USA  

$^6$ Max-Planck-Institut f\"{u}r Kernphysik, Saupfercheckweg 1, D-69117 Heidelberg, Germany } 

\end{center}

\rule{16.5cm}{0.3mm}

\bigskip

\footnotesize
{\bf Abstract} \hskip 5mm  Using a new approximate analytic parameter-free proxy-SU(3) scheme, we make predictions of shape observables for actinides and superheavy elements, namely $\beta$ and $\gamma$ deformation variables, 
and compare these with predictions by relativistic and non-relativistic mean-field theories. Furthermore, we make predictions  for $B(E2)$ transition rates of deformed nuclei and compare these with existing data and predictions of other theoretical approaches.
\bigskip


 \rule{16.5cm}{0.3mm}
\pagestyle{empty}

\normalsize

\bigskip\noindent
{\bf INTRODUCTION}

\medskip\noindent

The proxy-SU(3) is a new algebraic approach to heavy deformed nuclei based on fermionic symmetries, which has been introduced recently \cite{proxy1,proxy2}. Its basic assumptions and microscopic justification have been discussed in Ref. \cite{proxy1} and are further considered in the present Symposium  in Ref. \cite{Bonat}. A first success of the proxy-SU(3) scheme is the explanation of the prolate over oblate dominance in deformed nuclei, and the determination of the border of the prolate to oblate transition, 
which have been considered in Refs. \cite{proxy2,EPJA} and are further discussed in the present Symposium in Ref. \cite{Bonat}. In addition, parameter-free predictions for the deformation parameters $\beta$ and $\gamma$ for even rare earths have been made in Ref. \cite{proxy2} and successfully compared to Relativistic Mean Field predictions and to existing data. 

In the present work we first obtain parameter-free predictions for the deformation parameters $\beta$ and $\gamma$ for  even actinides and superheavy elements and we compare them to alternative theoretical predictions. Subsequently we calculate $B(E2)$ transition rates for deformed rare earth nuclei. 

\bigskip\noindent
{\bf EVEN-EVEN ACTINIDES AND SUPERHEAVY ELEMENTS} 
\medskip\noindent

For the illustrative and pedagogical purposes of this work, we take the relevant shells for the actinides and superheavy elements (SHE) as $Z = 82$-126,  $N = 126$-184, and $N=184$-258, although the upper bounds are by no means certain and microscopic calculations give many varying scenarios.
Parameter independent proxy-SU(3) results have been obtained for actinides and superheavy nuclei with $Z$ in the $pfh$ proxy-SU(3) shell (which is an approximation of the 82-126 shell) and neutrons in the sdgi proxy-SU(3) shell (which is an approximation of the 126-184 shell), 
as well as in the pfhj proxy-SU(3) shell (which is an approximation of the 184-258 shell), following 
the procedure described in Ref. \cite{proxy2}. Results are shown for $84\leq Z \leq 98$,
while additional results for $100\leq Z \leq 114$ have been presented elsewhere \cite{BonSDANCA}.
In order to have results from alternative calculations to compare our results with, 
we confine ourselves to $128\leq N\leq 220$.

We compare our results to predictions contained in the following sources:

1) Extended results for $10\leq Z\leq 110$ and $N\leq 200$ with the D1S Gogny interaction 
are given in \cite{Gogny} for the mean ground state $\beta$ deformation, as well as 
for the mean ground state $\gamma$ deformation.  

2) Extended results for $10\leq Z \leq 98$ and $16\leq N \leq 198$ by relativistic mean field theory (RMF-Lal) using the Lagrangian parametrization NL3 are reported in \cite{Lalazissis}. 

3) Extended results for the deformation $\beta$ up to $A=339$ are reported in the mass table FRDM(2012) \cite{Moller}, calculated within the finite-range droplet macroscopic model and the folded-Yukawa single-particle microscopic model. 

4) Extended results for the proton deformation $\beta_p$ and the neutron deformation 
$\beta_n$ with covariant density functional theory for $96\leq Z \leq 130$
and $N$ from the proton drip line up to $N=196$ are given in Ref. \cite{Ring} for the functionals PC-PK1 and DD-PC1. 

5) Extended results for the deformation $\beta$ within a microscopic-macroscopic method (MMM) for $98\leq Z \leq 126$ and $134\leq N \leq 192$ are given in Ref. \cite{Skalski}. 

The results are summarized in Figs. 1 and 2. Overall good agreement is observed between the 
parameter-free proxy-SU(3) predictions and the alternative calculations. 

\bigskip\noindent
{\bf B(E2) RATIOS}
\medskip\noindent

As discussed in Appendix A of Ref. \cite{MartSDANCA}, $B(E2)$s within the proxy-SU(3) model are proportional
to the square of the relevant reduced matrix element of the quadrupole
operator $Q$. If ratios of $B(E2)$s within the same nucleus and within the same irreducible representation are considered, only
the relevant SU(3)$\to$SO(3) coupling coefficients remain, while all other factors
cancel out, leading to
\begin{equation} 
{B(E2;L_i\to L_f)\over B(E2;2_g\to 0_g)} = 5 {2L_f+1\over 2L_i+1} \\
{( \langle (\lambda,\mu) K_i L_i ; (1,1)2 || (\lambda,\mu) K_f L_f \rangle)^2 \over 
( \langle (\lambda,\mu) 0 2 ; (1,1)2 || (\lambda,\mu) 0 0 \rangle)^2}, \label{BE2}
\end{equation} 
where normalization to the $B(E2)$ connecting the first excited $2^+$ state to the
$0^+$ ground state of even-even nuclei is made. The needed SU(3)$\to$SO(3)
coupling coefficients are readily obtained from the SU3CGVCS code \cite{RBcode}, as
described in Appendix A of Ref. \cite{MartSDANCA}. 

It should be noticed that the ratios given by Eq. (\ref{BE2}) are completely free of any
free parameters and/or scaling factors.

\bigskip\noindent
{\bf NUMERICAL RESULTS}
\medskip\noindent

Calculations have been performed for the proxy-SU(3) irreps (54,12) and (52,14).
 The irrep (54,12) accommodates $^{168}$Er, for which complete spectroscopy
has been performed \cite{Warner},while the irrep (52,14) accommodates $^{162}$Dy, for which complete spectroscopy has been performed \cite{Aprahamian}, and $^{166}$Er, for which rich 
data exist \cite{ENSDF}.

$B(E2)$s within the ground state band are shown in the upper four panels of Fig. 3. Agreement between the
proxy-SU(3) predictions and the data is excellent in the cases of
$^{162}$Dy and $^{166}$Er, while in $^{168}$Er three points are missed. In addition, very good agreement is seen
in most cases with various predictions within the pseudo-SU(3) scheme.  It appears that nuclear stretching \cite{Pin} has been properly taken into account.

Proxy-SU(3) predictions for $B(E2)$s within the $\gamma_1$ band, with $\Delta L=-2$ (increasing
with $L$) and $\Delta L=-1$ (decreasing with $L$), are shown in the lower four panels of Fig. 3, and
are compared to the data for nuclei for which sufficient data exist \cite{ENSDF}, as well as to pseudo-SU(3) predictions.  The distinction
between increasing $B(E2)$s with $\Delta L = -2$ and decreasing $B(E2)$s with 
$\Delta L = -1$ is seen clearly in the data. Agreement with the pseudo-SU(3) predictions is excellent in almost all cases. 

\bigskip\noindent
{\bf DISCUSSION}
\medskip\noindent

The main findings of the present section can be summarized as follows.

Analytic expressions for $B(E2)$ ratios for heavy deformed nuclei providing
numerical results in good agreement with experiment are derived within
the proxy-SU(3) scheme without using any free parameters
and/or scaling factors. The derivation, described in Appendix A of Ref. \cite{MartSDANCA}, is exact.  The only quantities appearing in the final formula are the relevant
SU(3)$\to$SO(3) coupling coefficients, for which computer codes are readily available
\cite{RBcode,Akicode}.

Concerning further work, spectra of heavy deformed nuclei will be considered
within the proxy-SU(3) scheme, involving three- and/or four-body terms in order
to break the degeneracy between the ground state and $\gamma_1$ bands  \cite{DW2,DW1,PVI}. Furthermore,
$B(M1)$ transition rates can be considered along the proxy-SU(3) path, using
the techniques already developed \cite{CDL} in the framework of the pseudo-SU(3)
scheme.

\bigskip\noindent
{\bf ACKNOWLEDGEMENTS}
\medskip\noindent

Work partly supported by the Bulgarian National Science Fund (BNSF) under Contract No. DFNI-E02/6, by the MSU-FRIB laboratory, by the Max Planck Partner group, TUBA-GEBIP, and by the Istanbul University Scientific Research Project No. 54135.


\begin{figure*}[htb]

{\includegraphics[width=70mm]{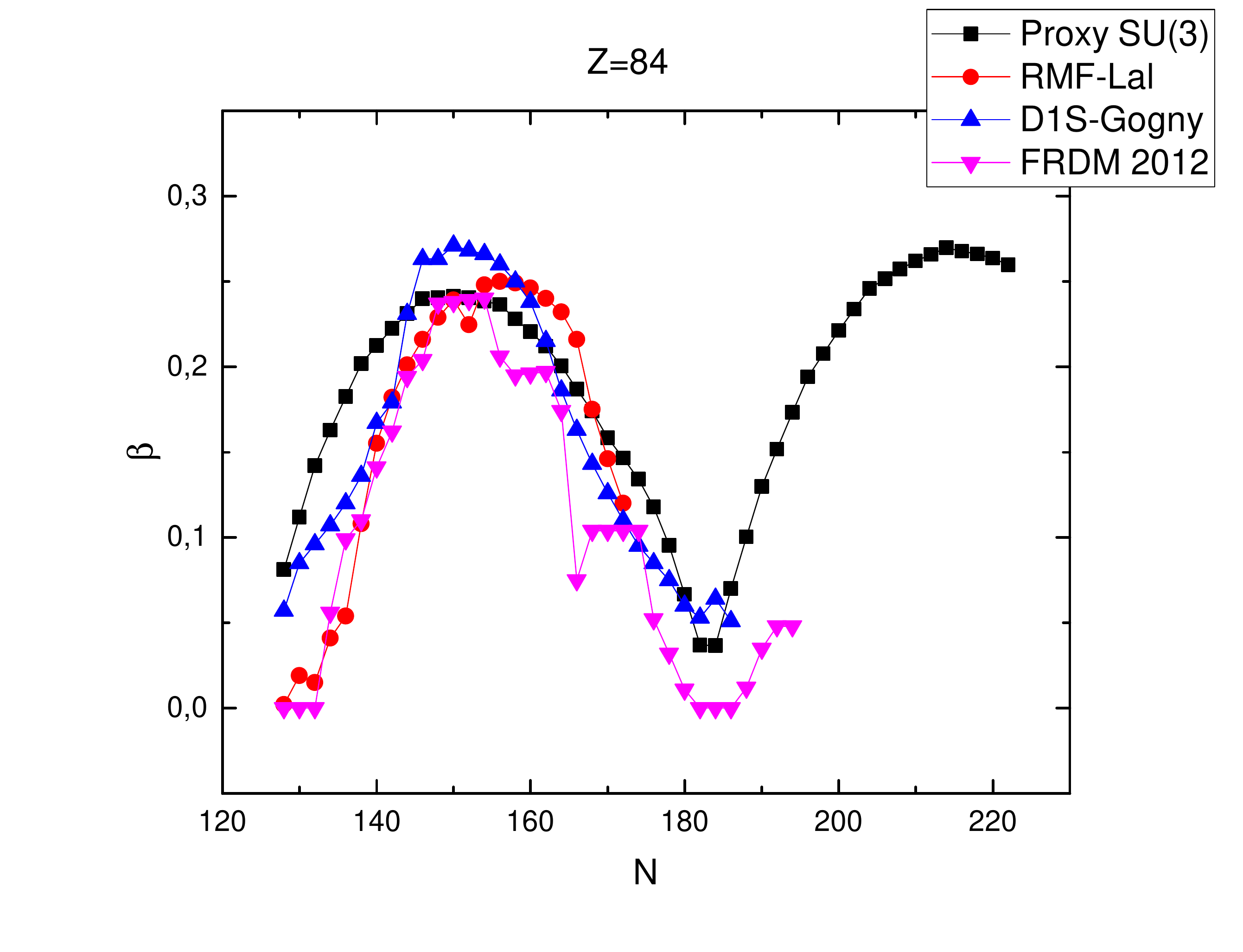}\hspace{5mm}
\includegraphics[width=70mm]{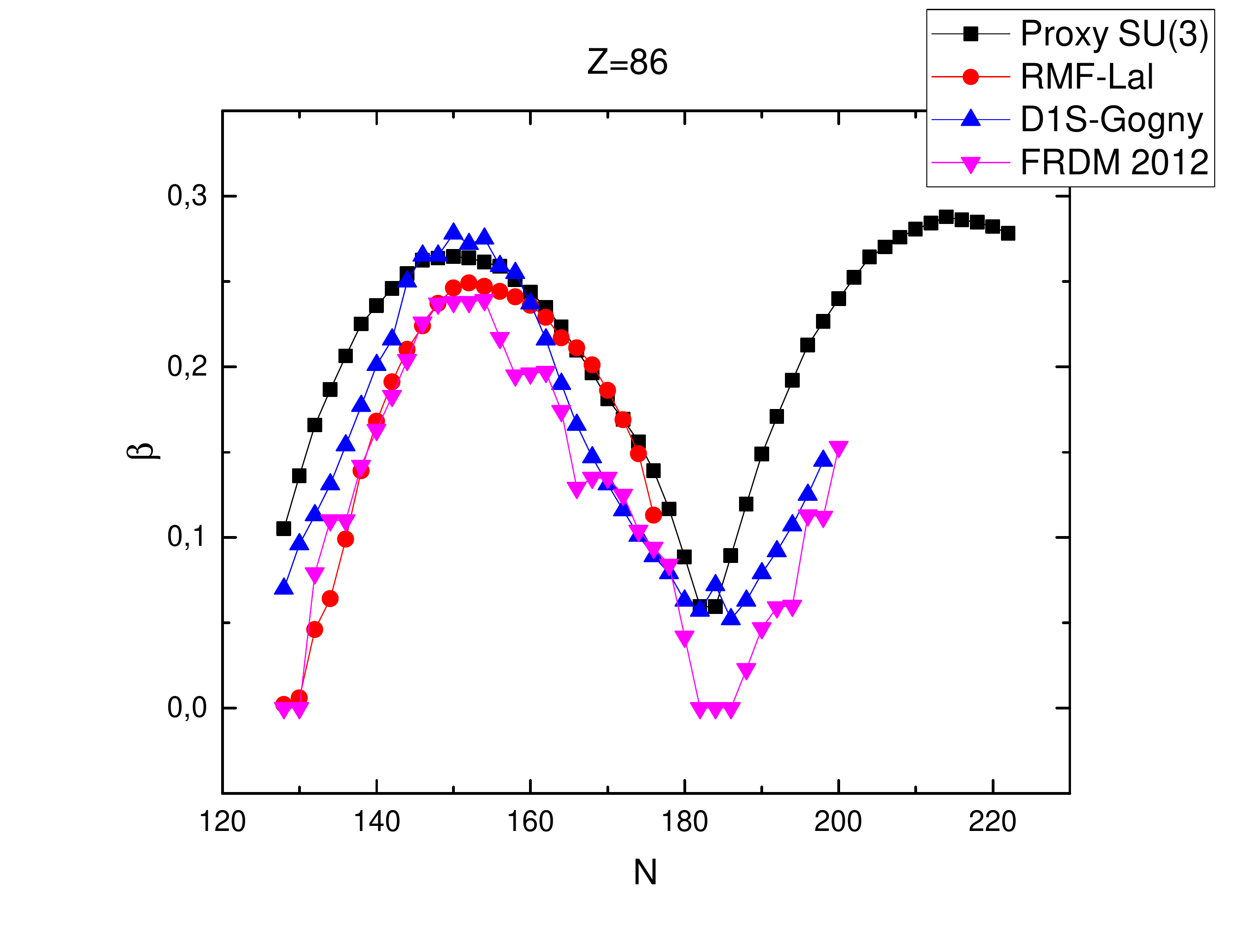}}
{\includegraphics[width=70mm]{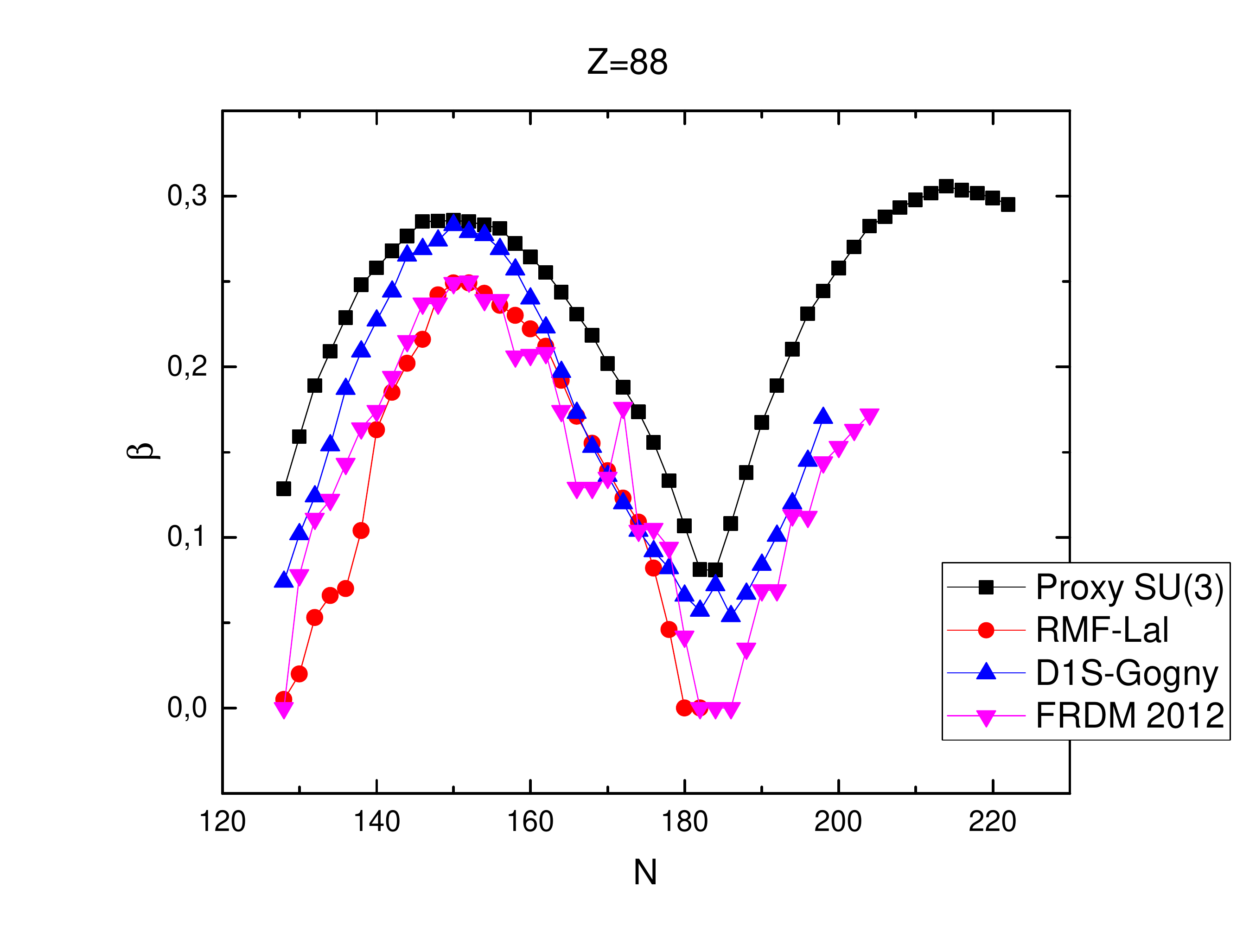}\hspace{5mm}
\includegraphics[width=70mm]{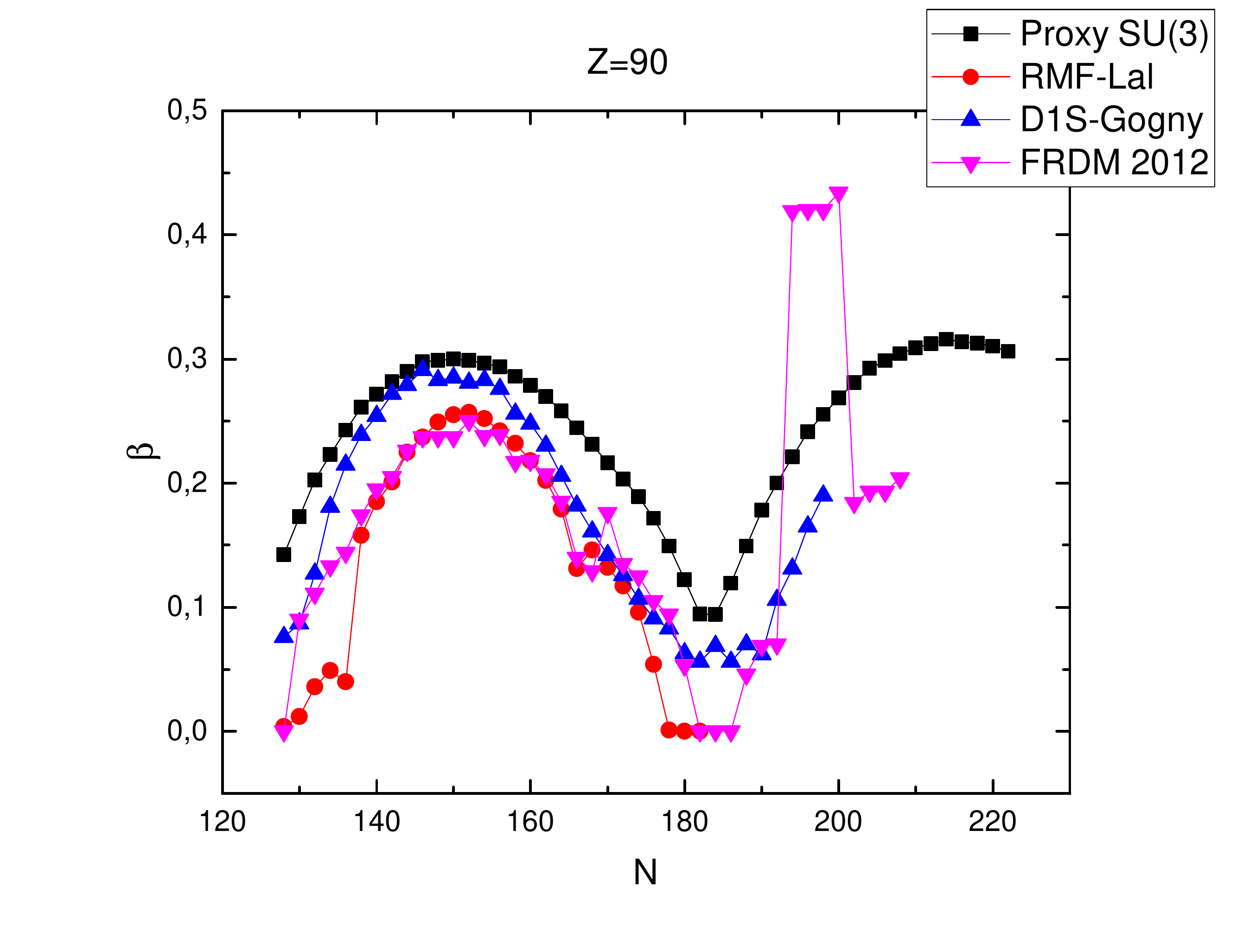}}
{\includegraphics[width=70mm]{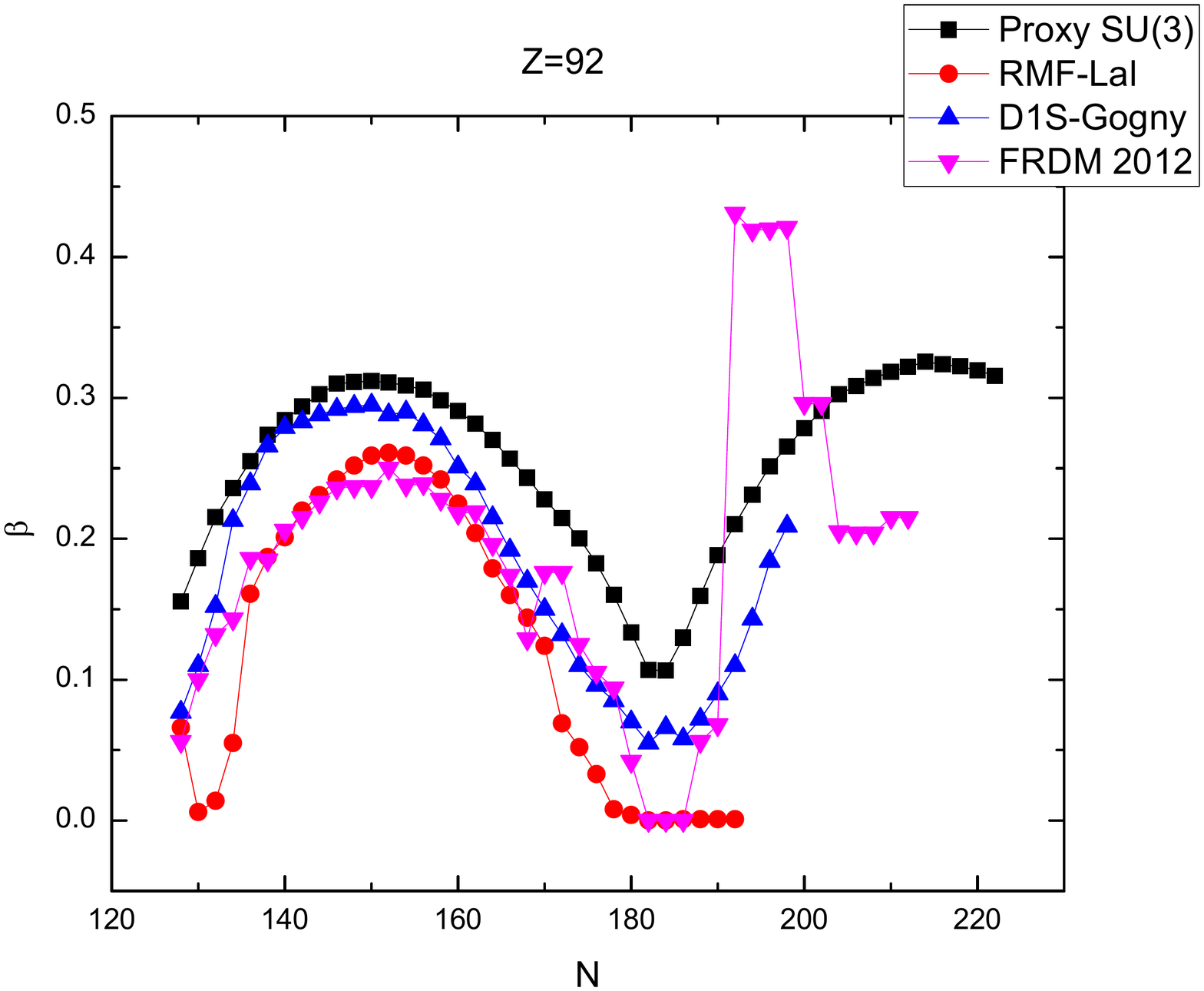}\hspace{5mm}
\includegraphics[width=70mm]{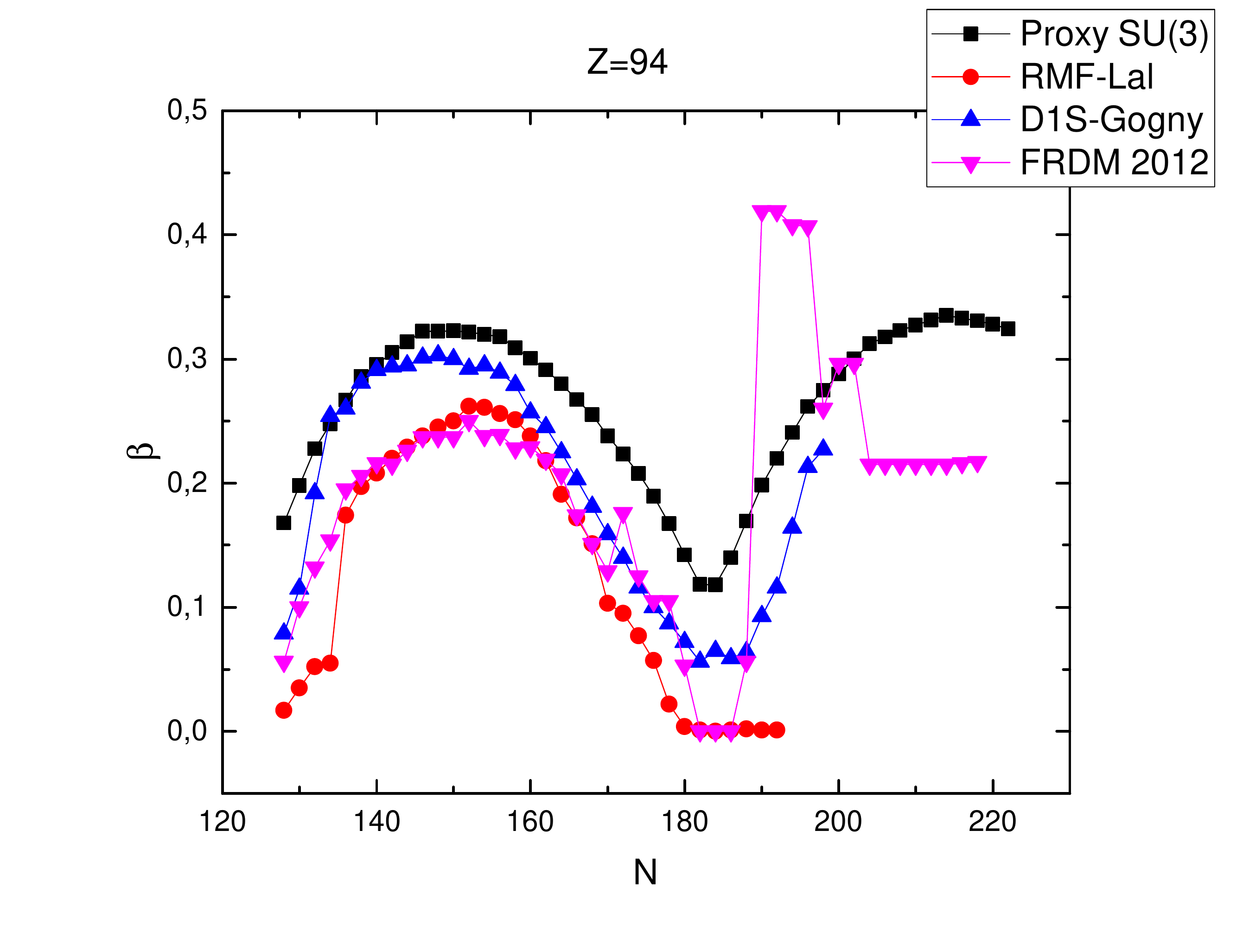}}
{\includegraphics[width=70mm]{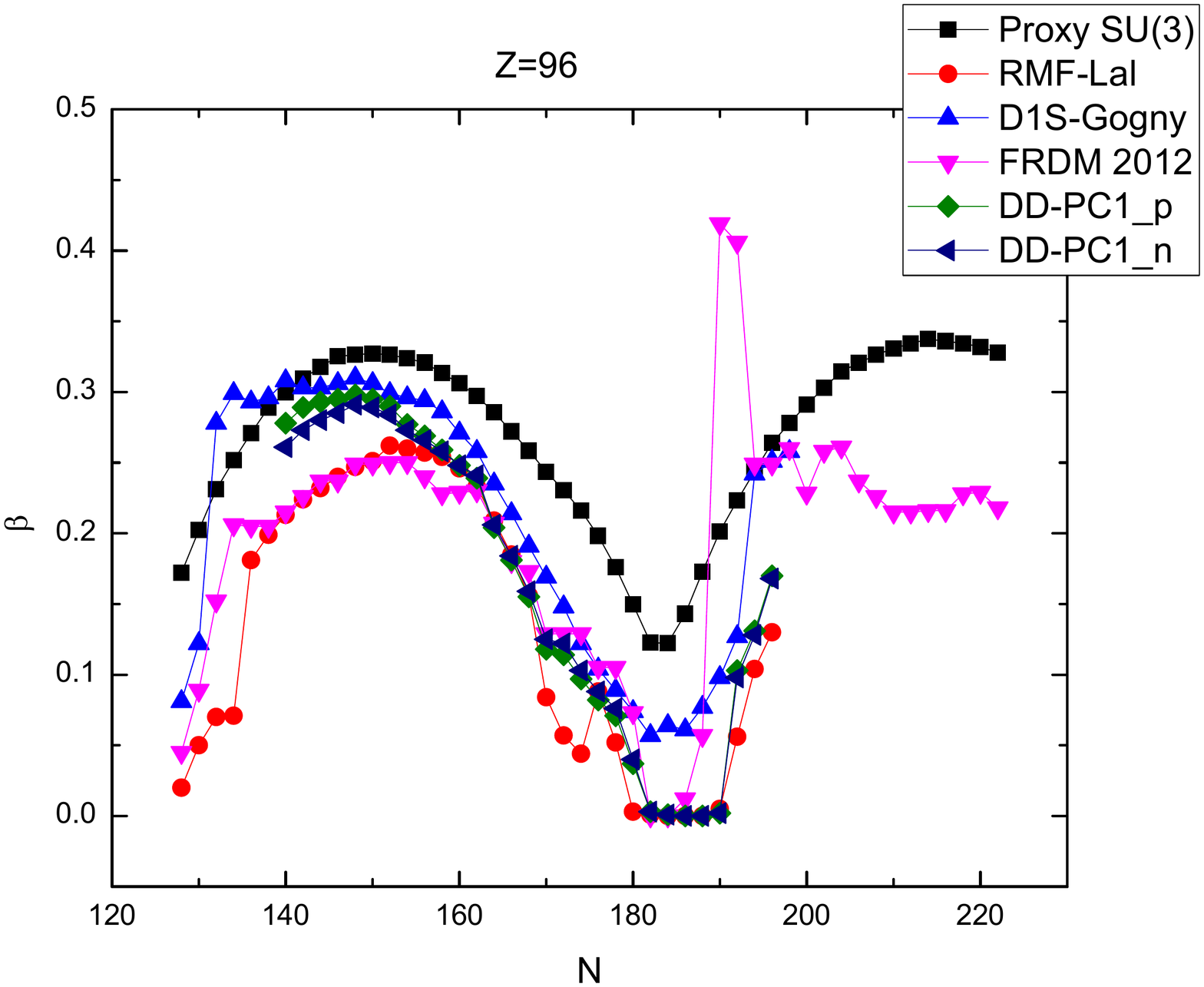}\hspace{35mm}
\includegraphics[width=70mm]{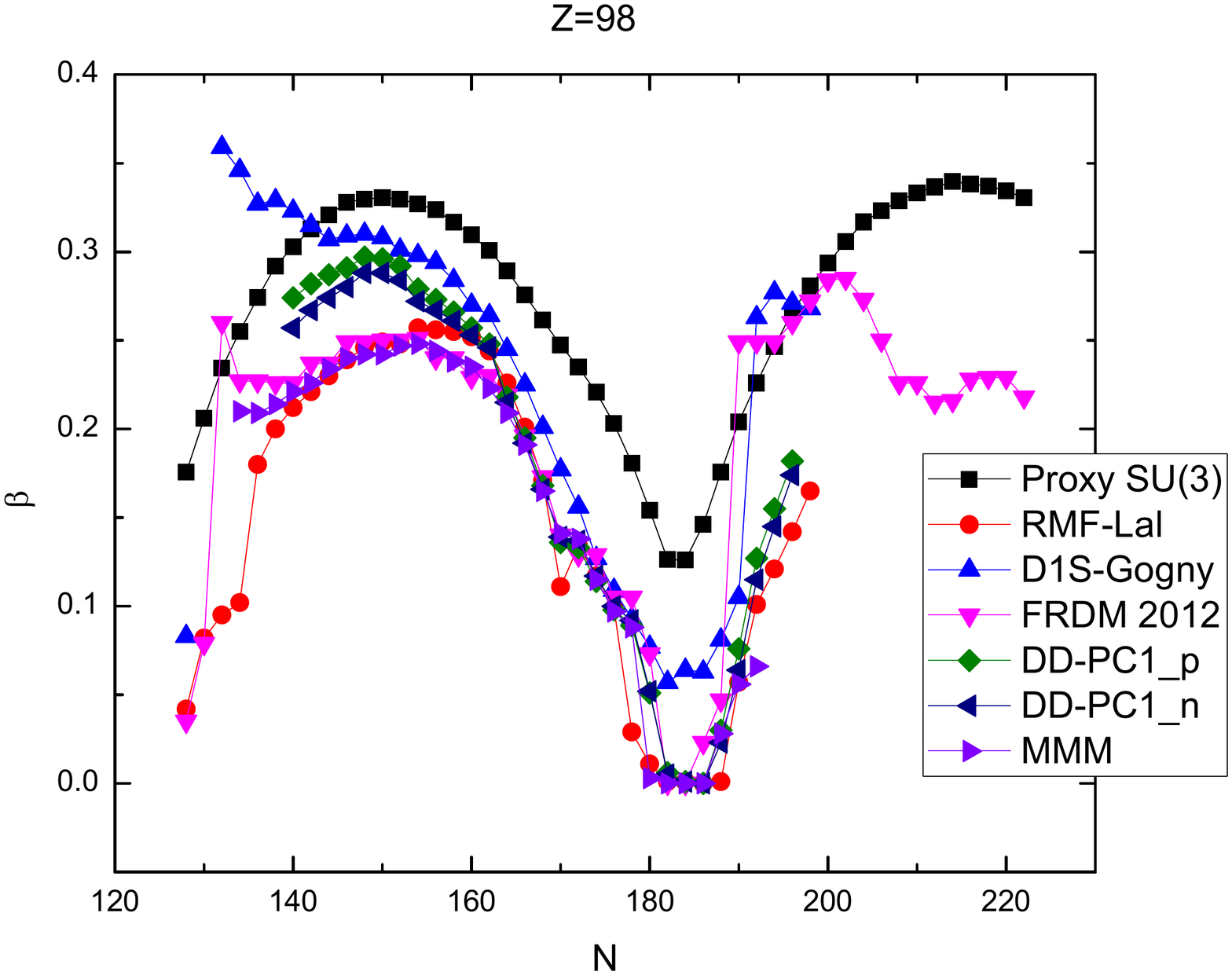}}

\caption{\footnotesize Proxy SU(3) predictions for $Z=84$-98 for $\beta$, obtained following the procedure of Ref. \cite{proxy2}, 
compared with
results by relativistic mean field theory (RMF-Lal) \cite{Lalazissis}, the microscopic-macroscopic method (MMM) \cite{Skalski}, covariant density functional theory with 
the DD-PC1 functional (DD-PC1) \cite{Ring} (in which case different values for protons (DD-PC1\_p) and neutrons (DD-PC1\_n) are reported), the D1S-Gogny interaction (D1S-Gogny) 
\cite{Gogny}, and the mass table FRDM(2012)\cite{Moller}. } 

\end{figure*}


\begin{figure*}[htb]

{\includegraphics[width=70mm]{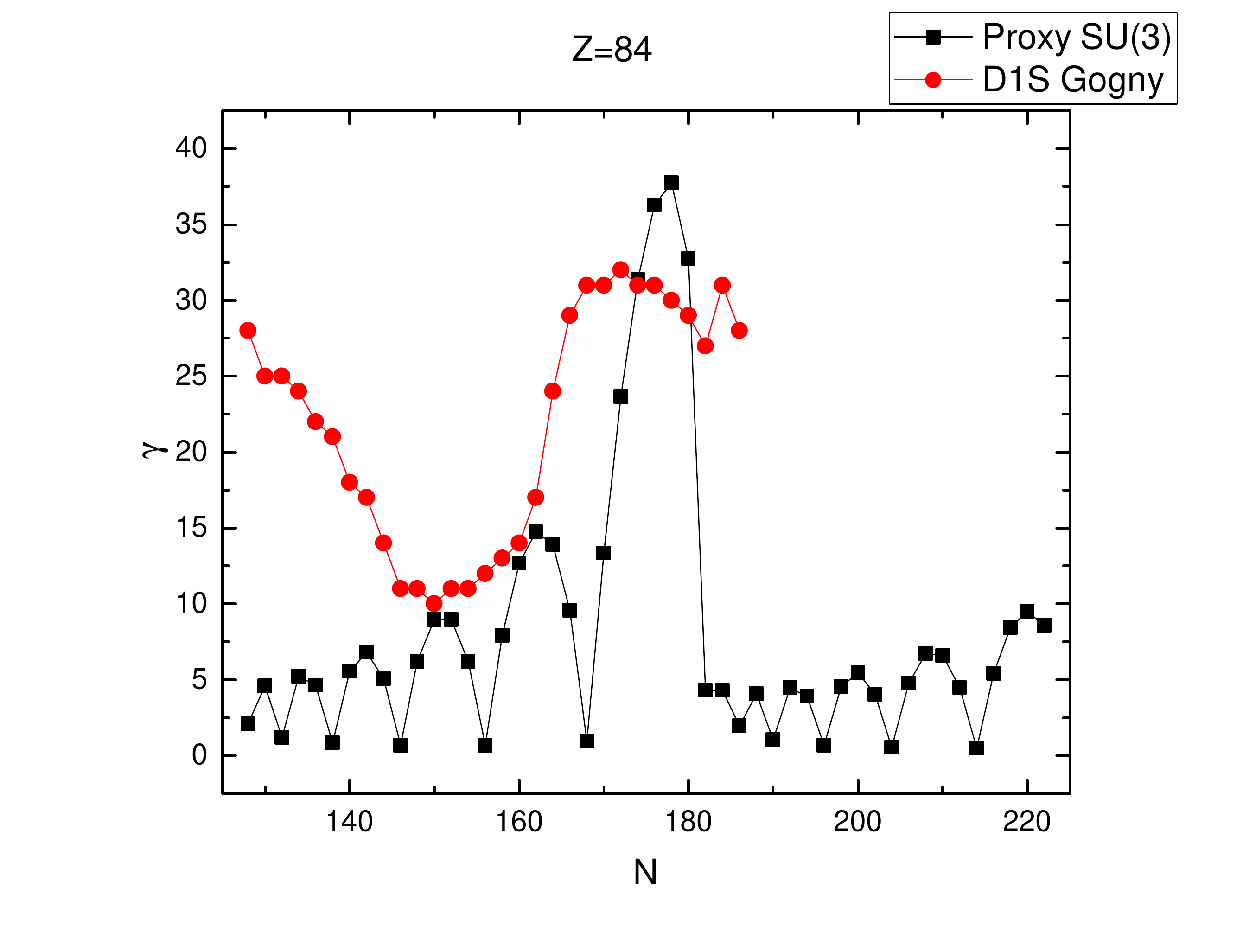}\hspace{5mm}
\includegraphics[width=70mm]{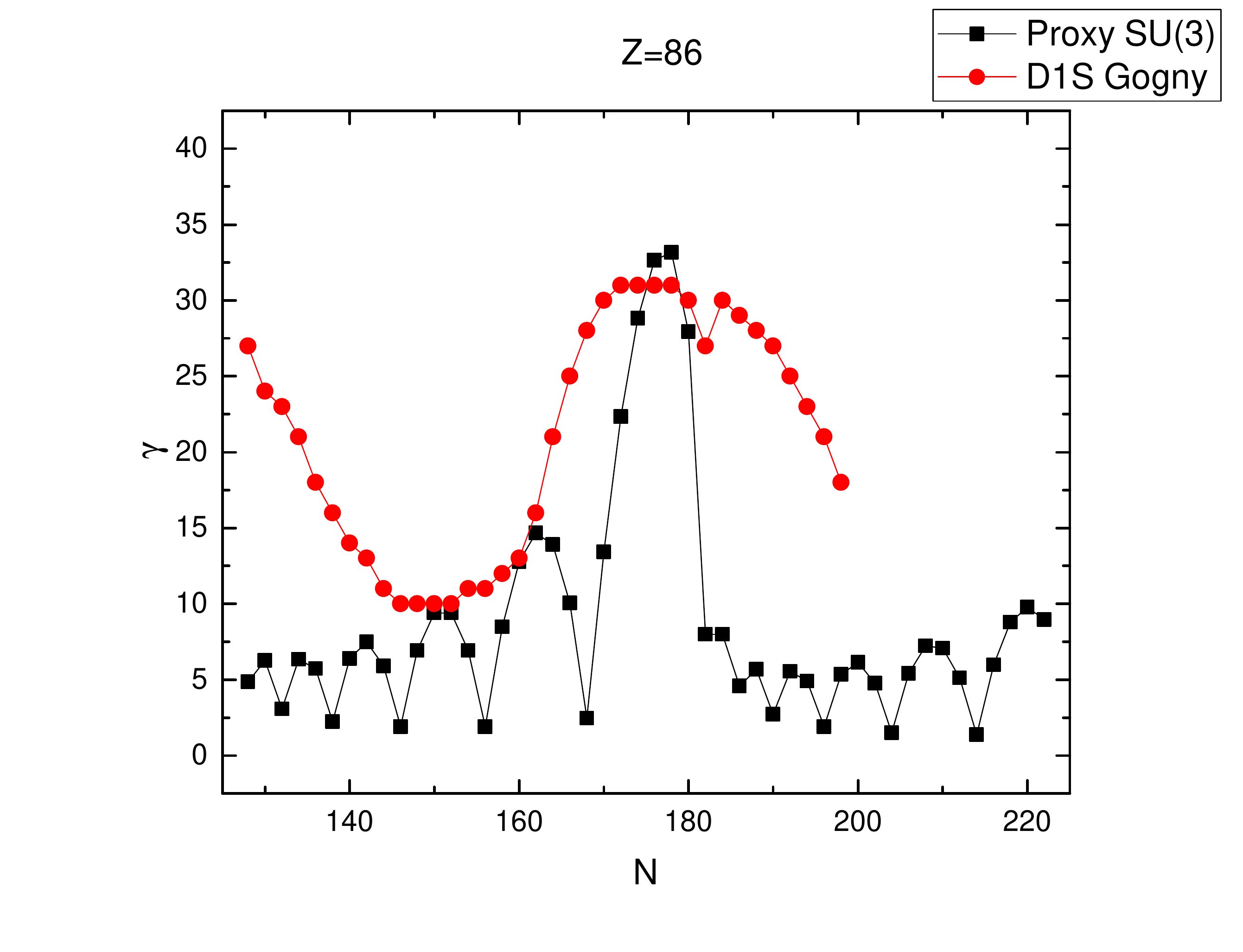}}
{\includegraphics[width=70mm]{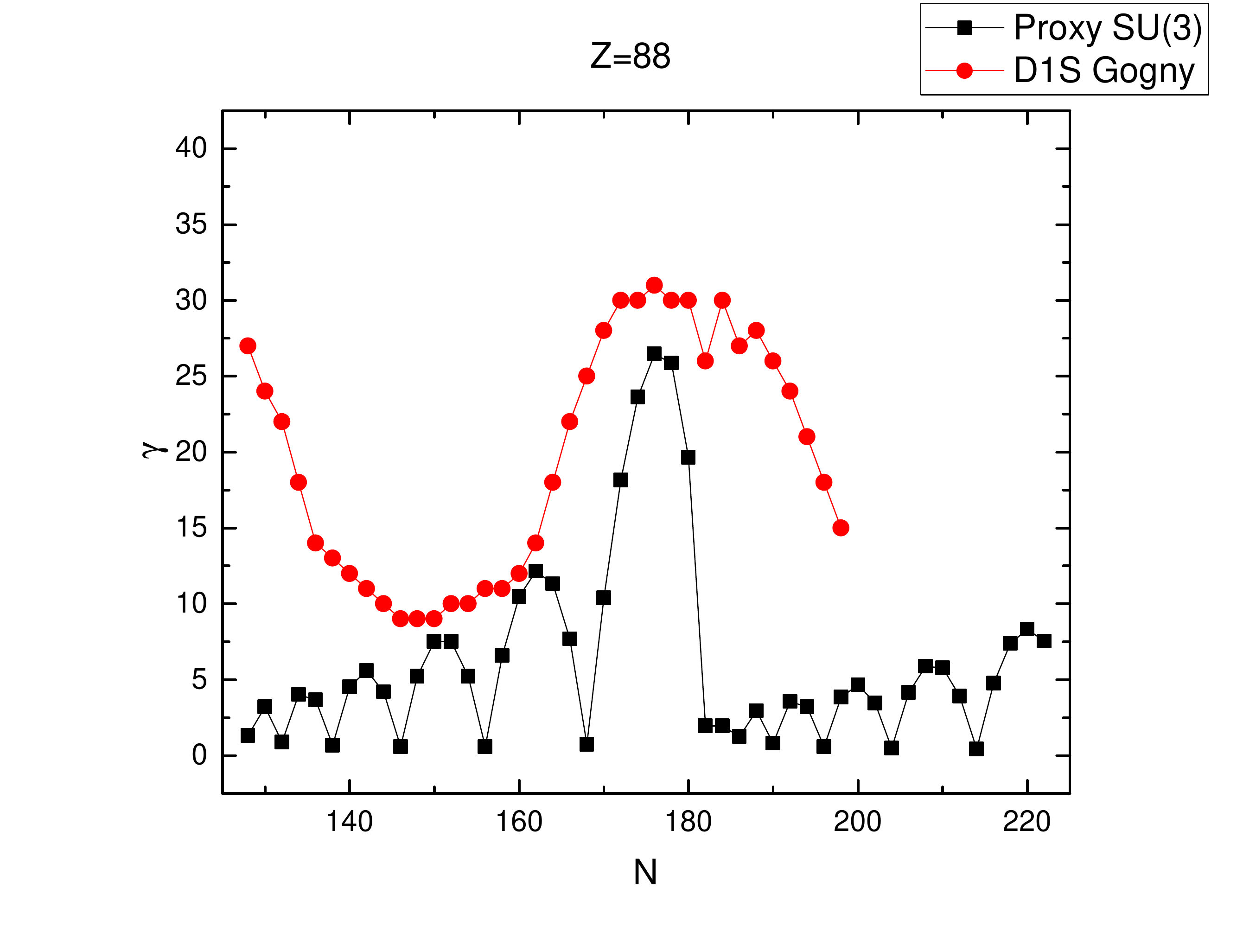}\hspace{5mm}
\includegraphics[width=70mm]{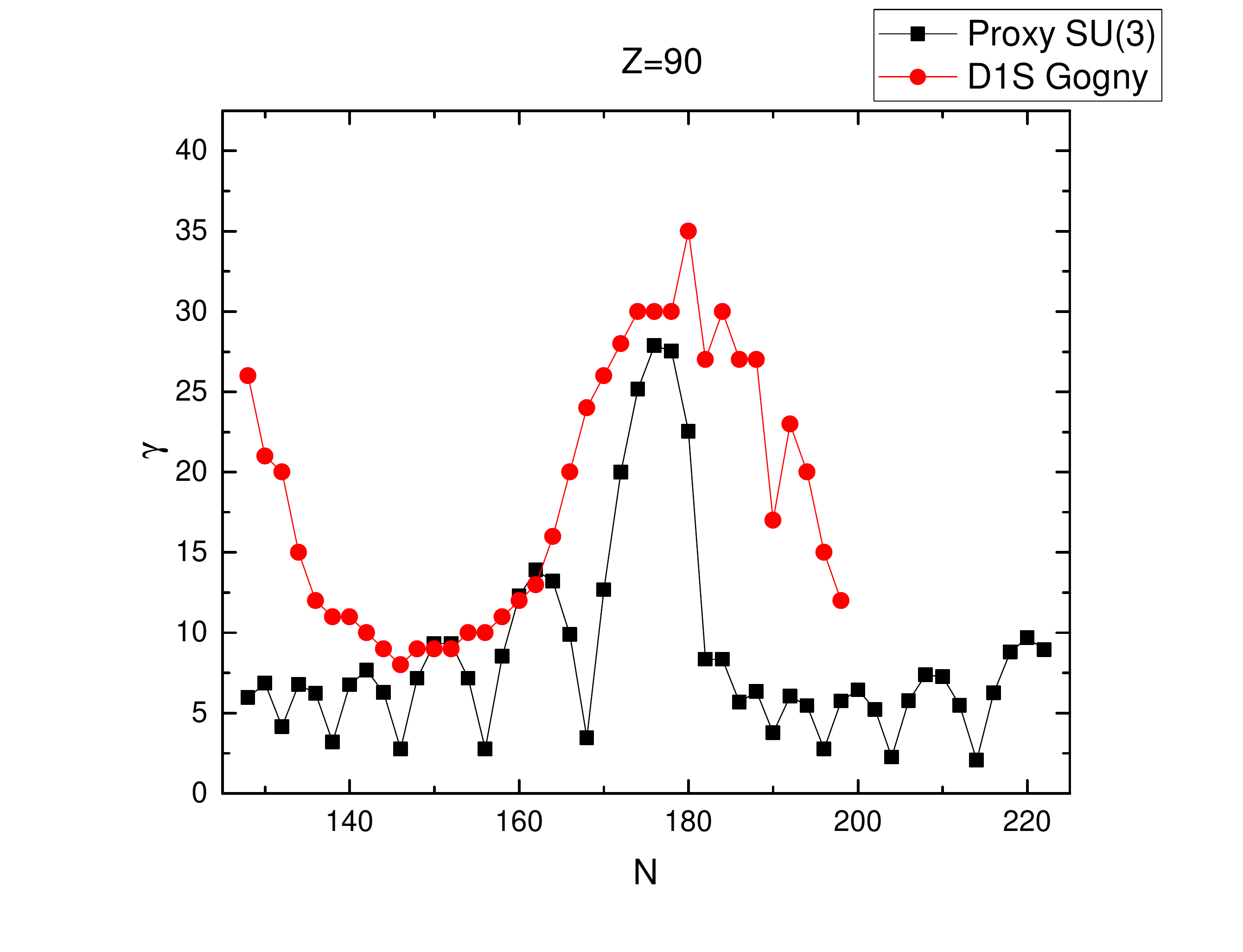}}
{\includegraphics[width=70mm]{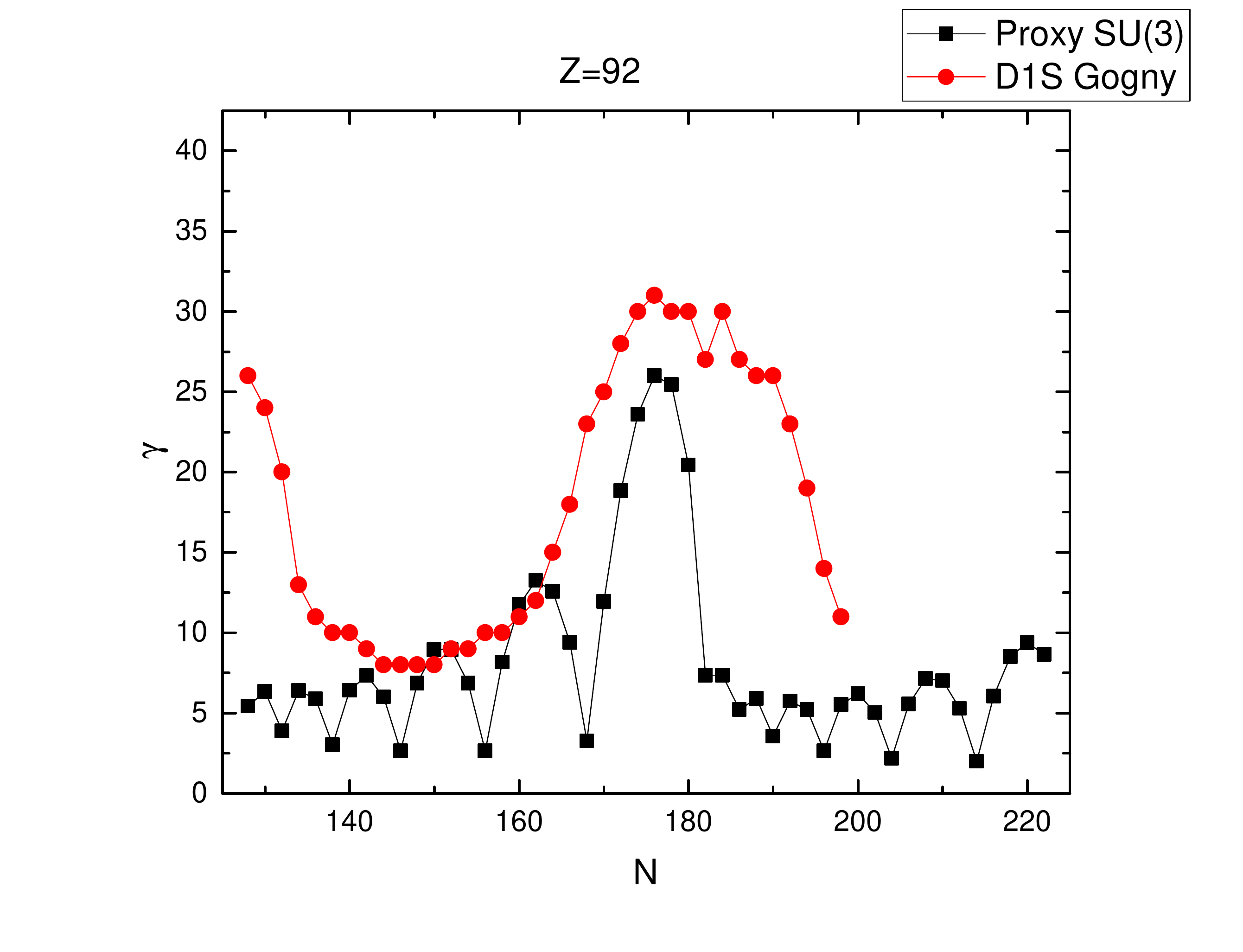}\hspace{5mm}
\includegraphics[width=70mm]{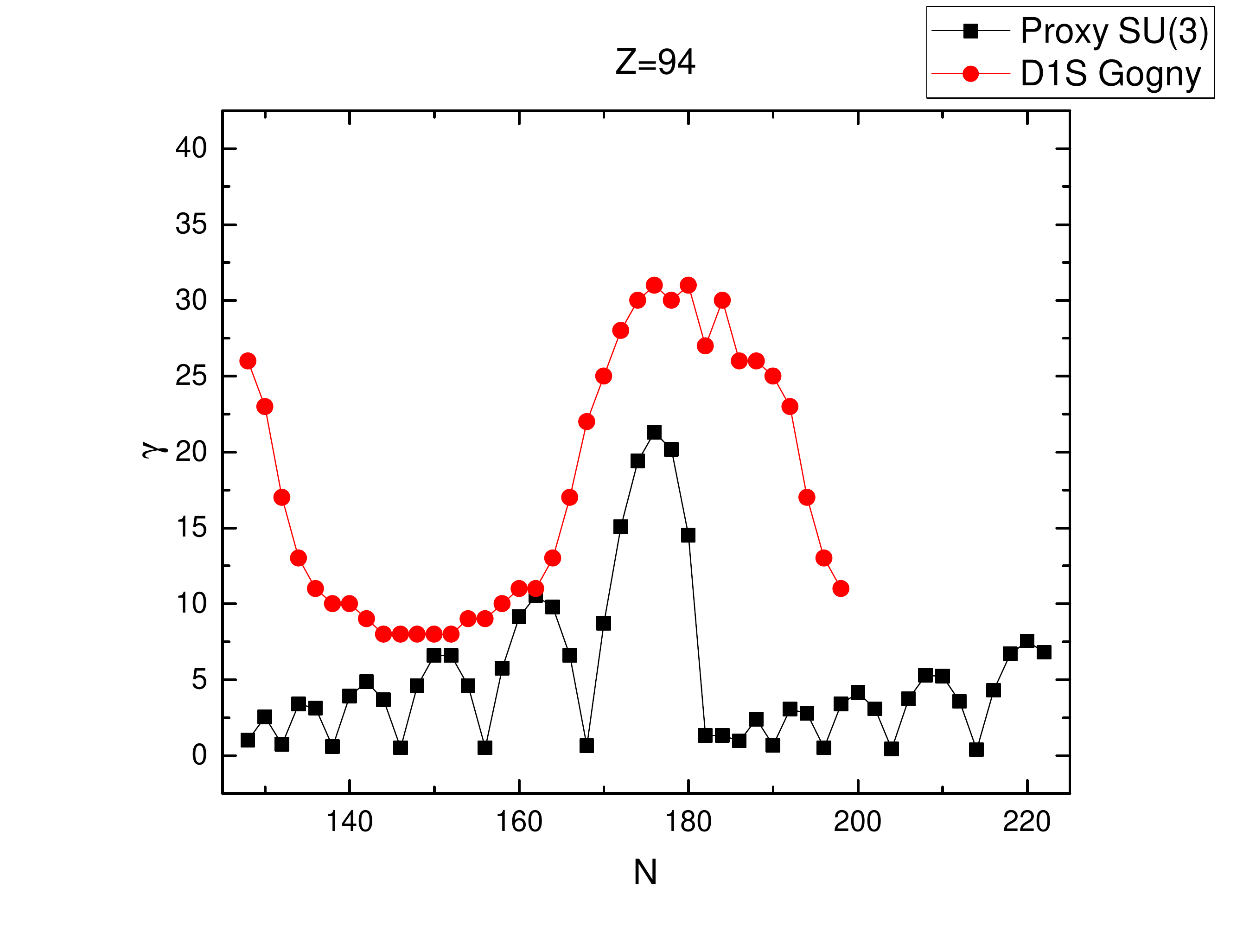}}
{\includegraphics[width=70mm]{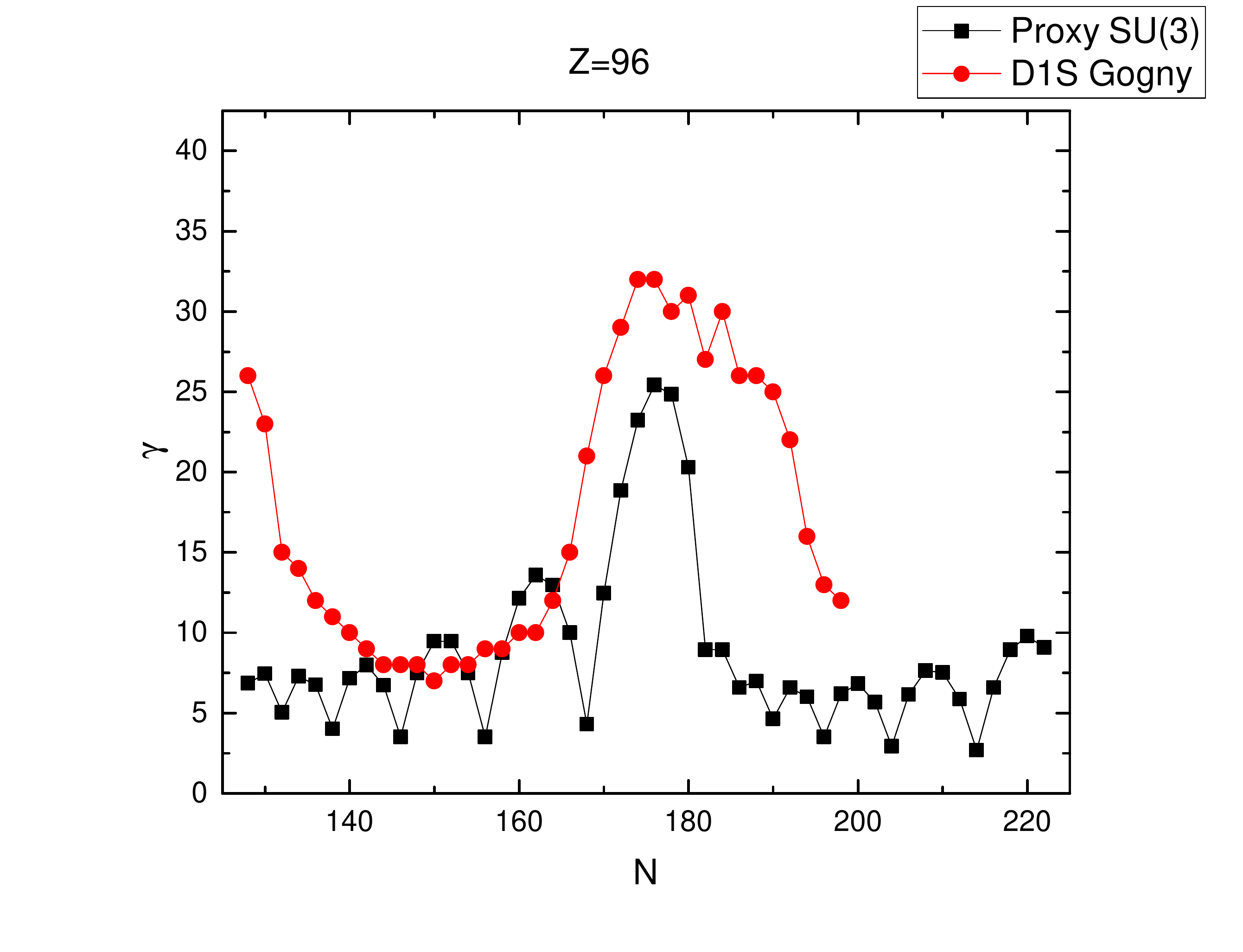}\hspace{35mm}
\includegraphics[width=70mm]{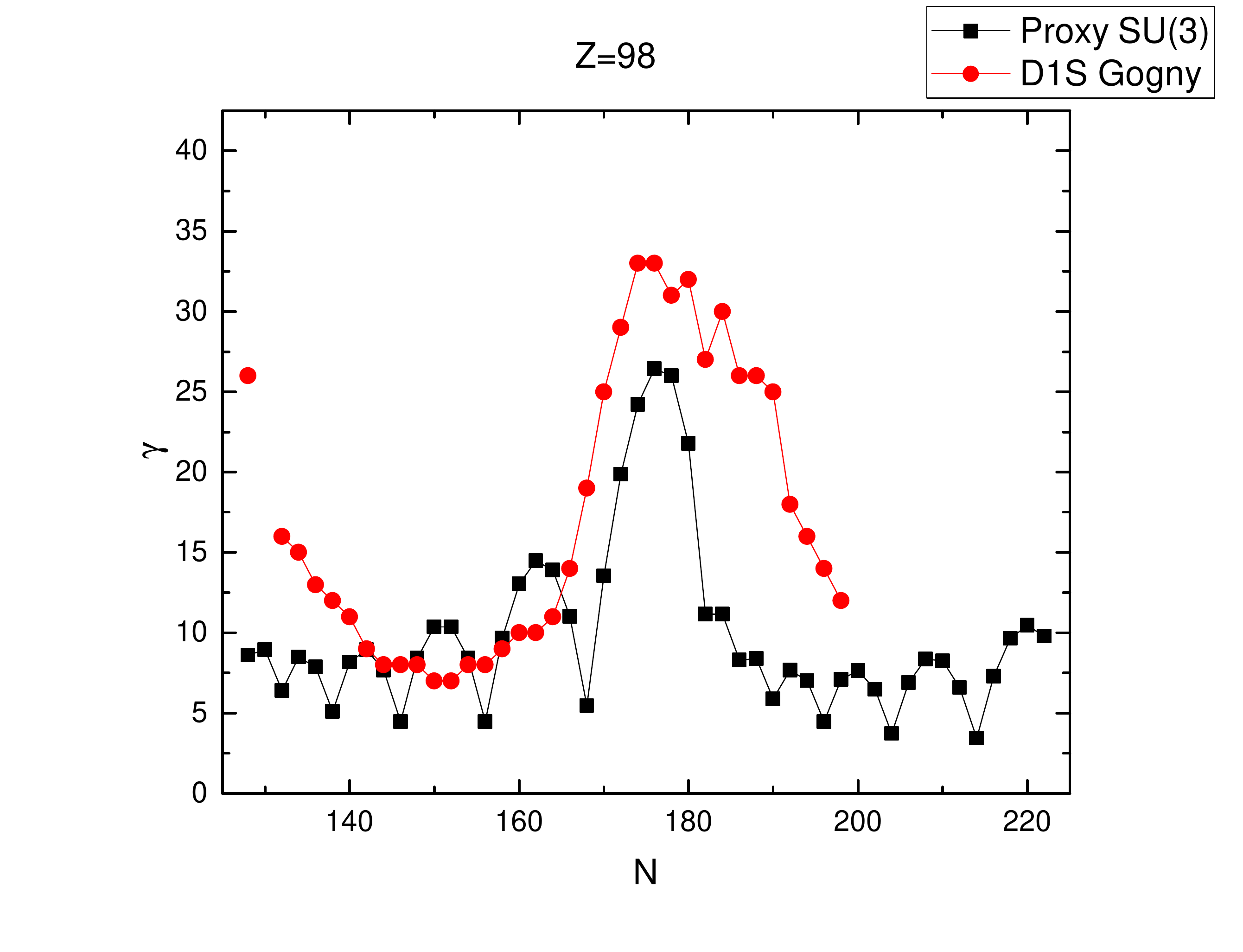}}

\caption{\footnotesize  Proxy SU(3) predictions for $Z=84$-98 for $\gamma$, obtained following the procedure of Ref. \cite{proxy2}, compared with results by the D1S-Gogny interaction (D1S-Gogny) \cite{Gogny}. } 

\end{figure*}


\begin{figure*}[htb]

{\includegraphics[width=70mm]{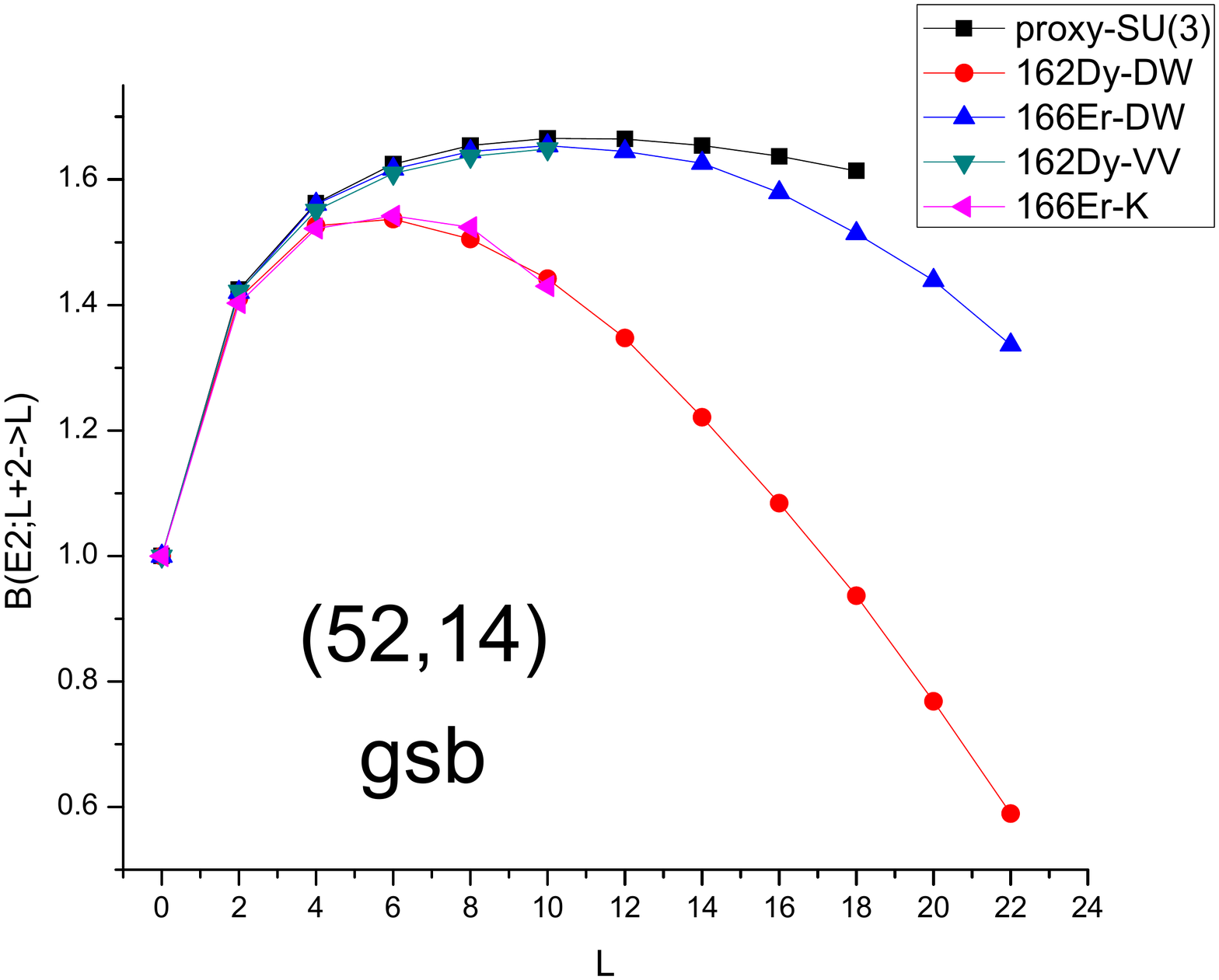}\hspace{5mm}
\includegraphics[width=70mm]{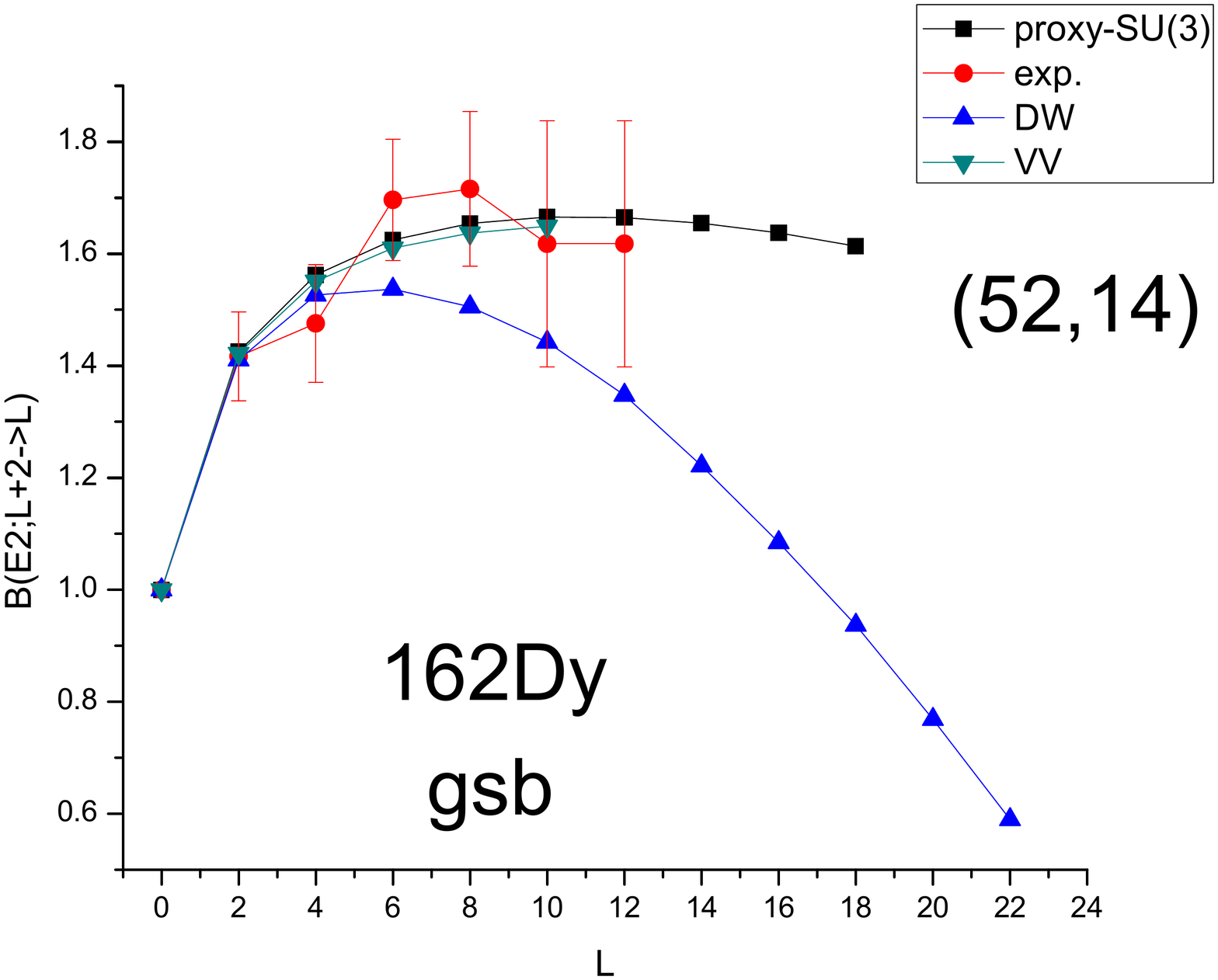}}
{\includegraphics[width=70mm]{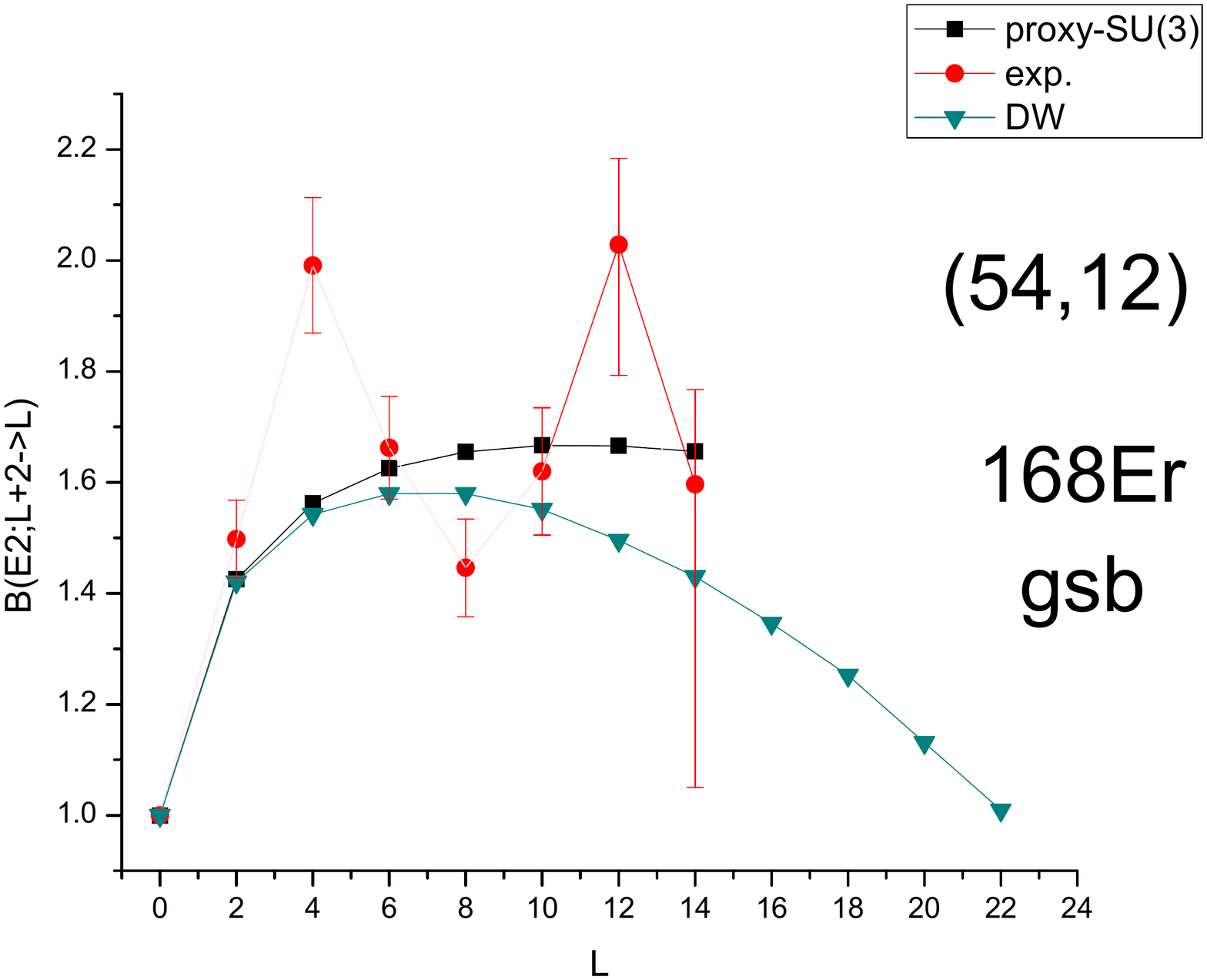}\hspace{5mm}
\includegraphics[width=70mm]{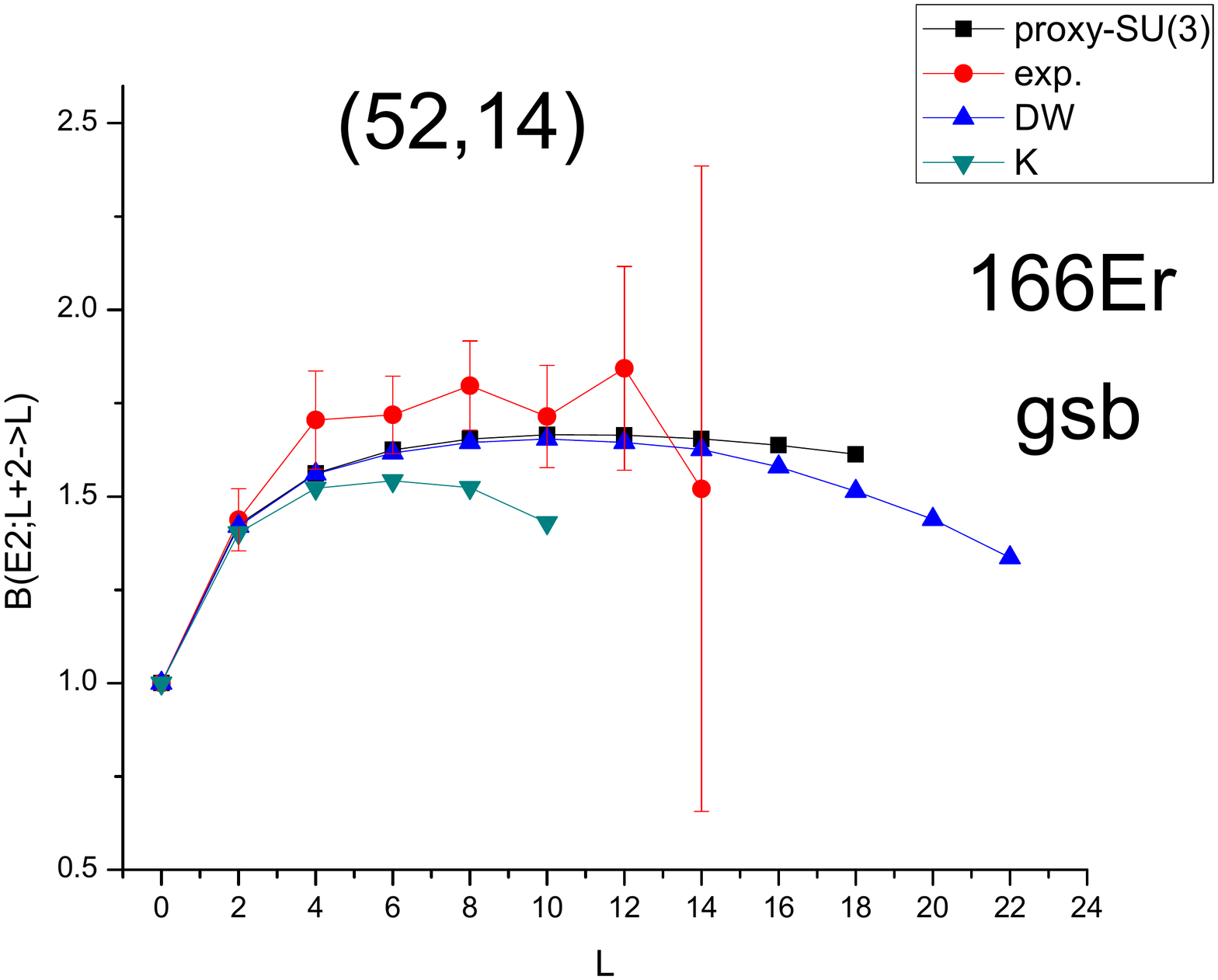}}
{\includegraphics[width=70mm]{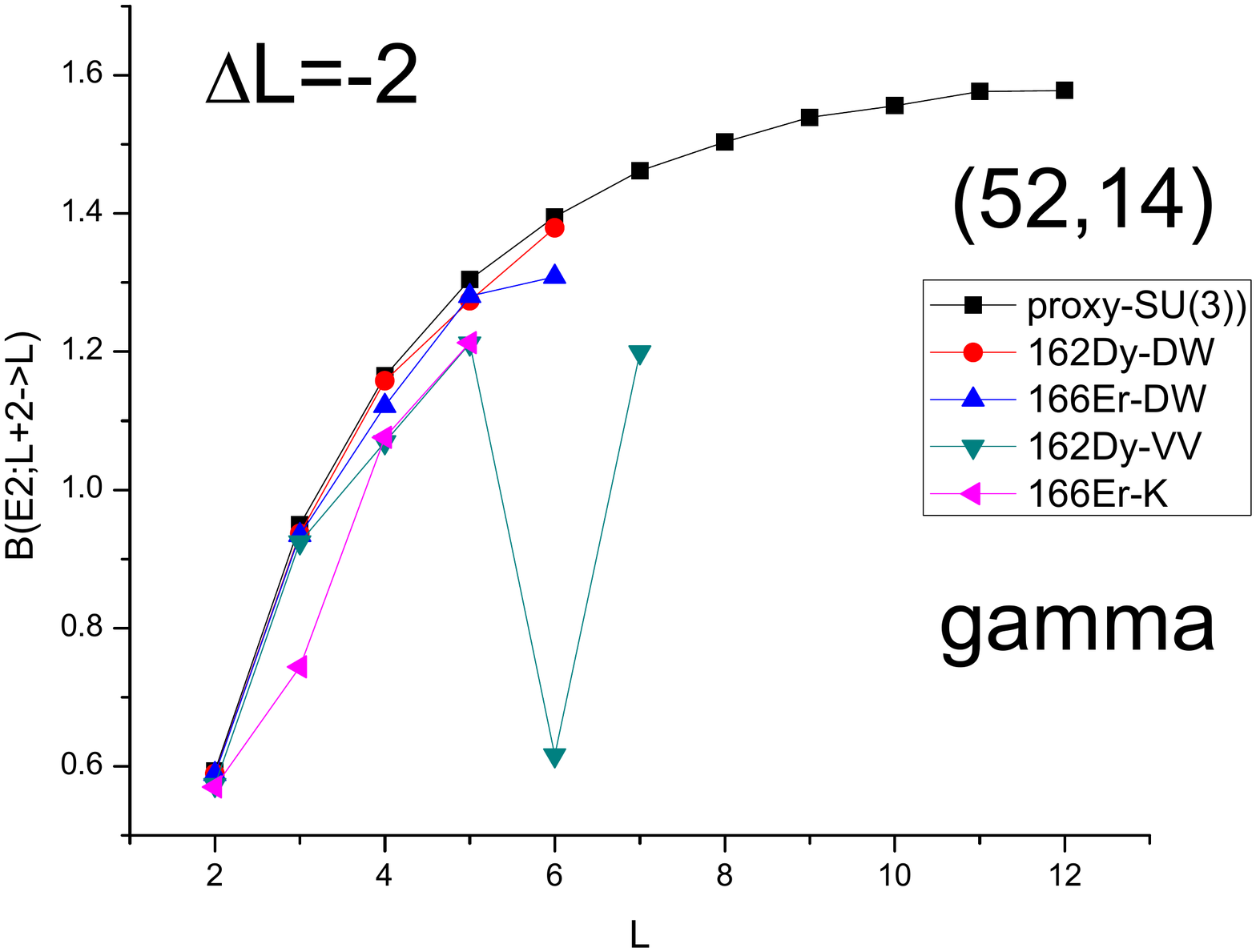}\hspace{5mm}
\includegraphics[width=70mm]{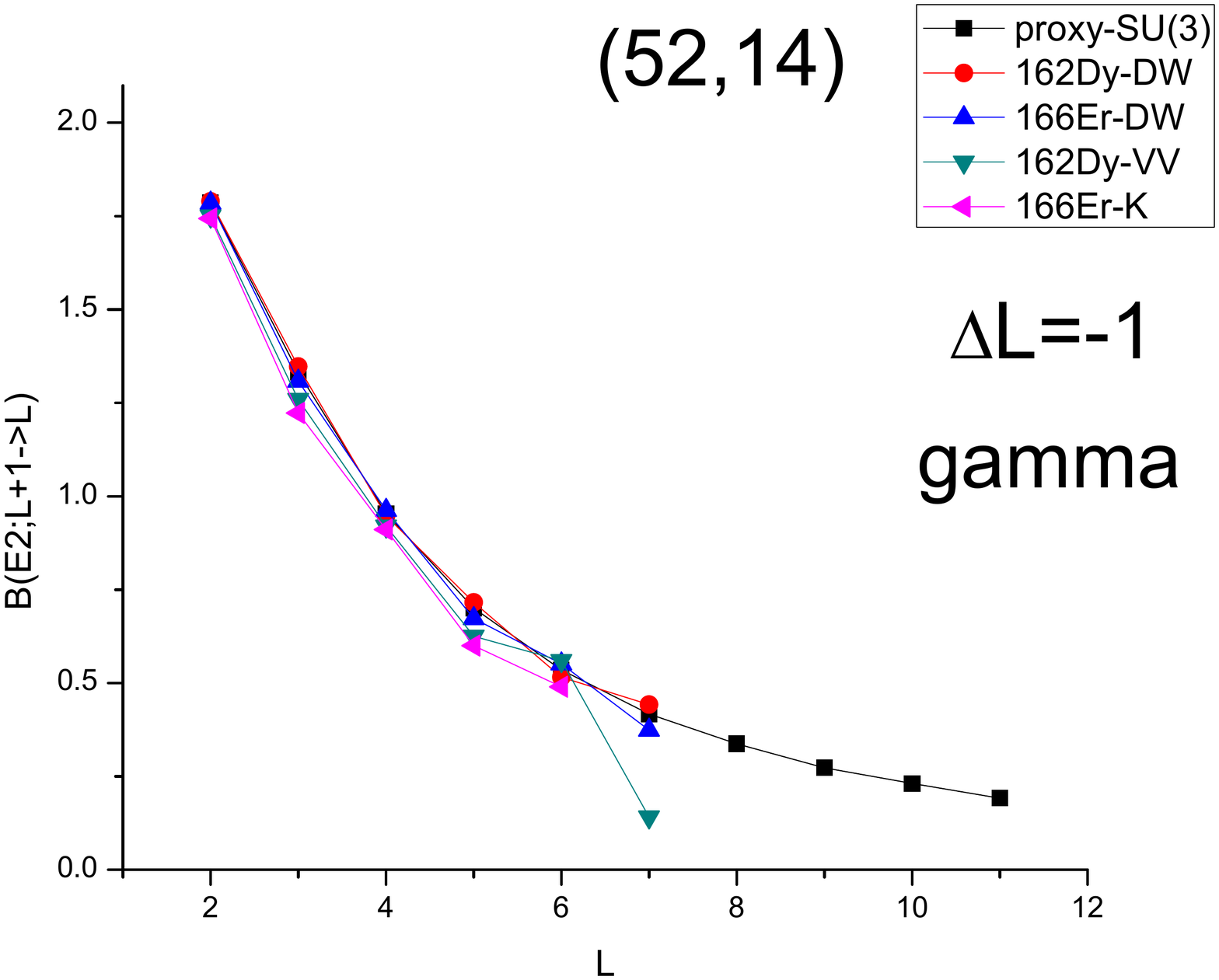}}
{\includegraphics[width=70mm]{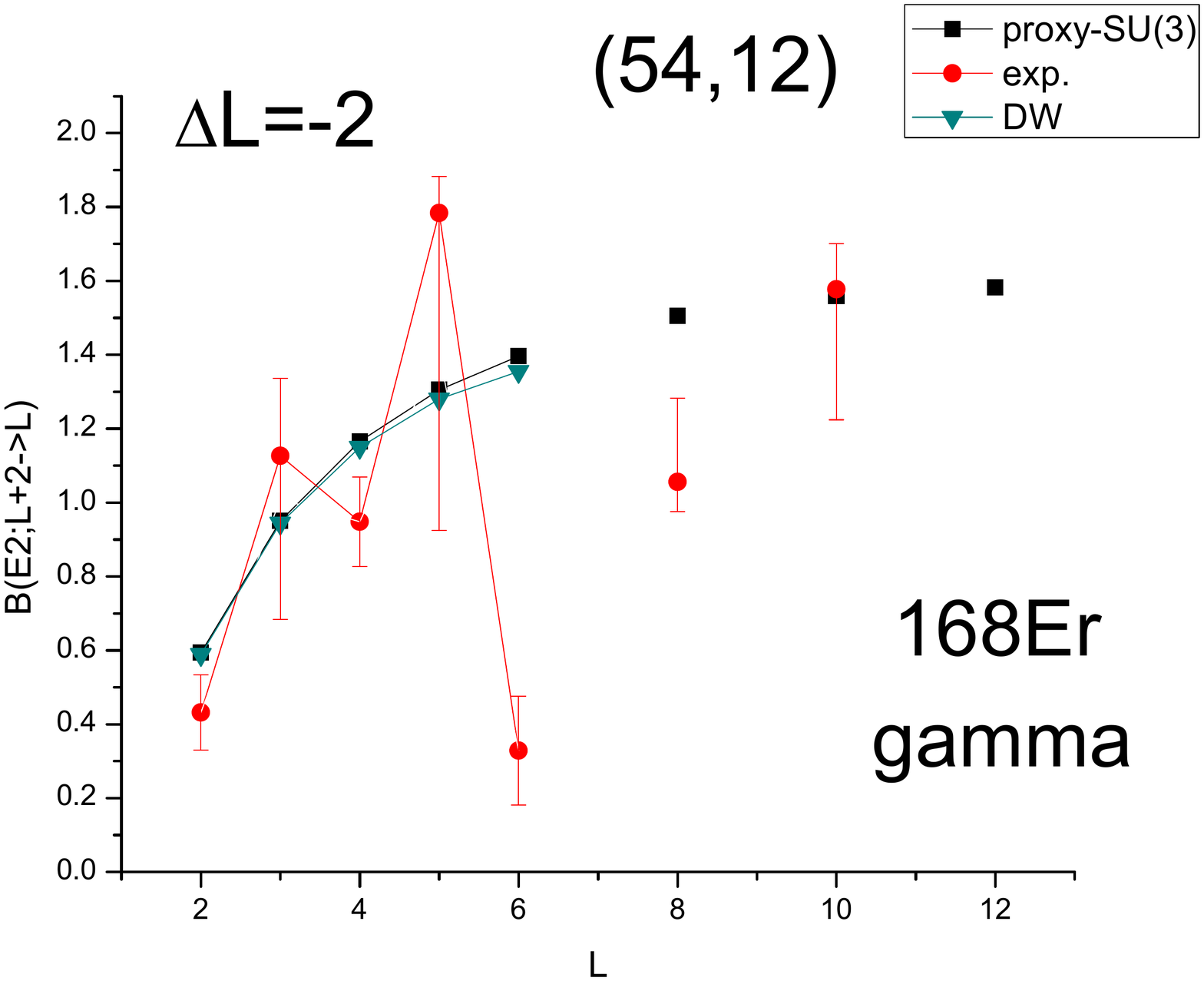}\hspace{35mm}
\includegraphics[width=70mm]{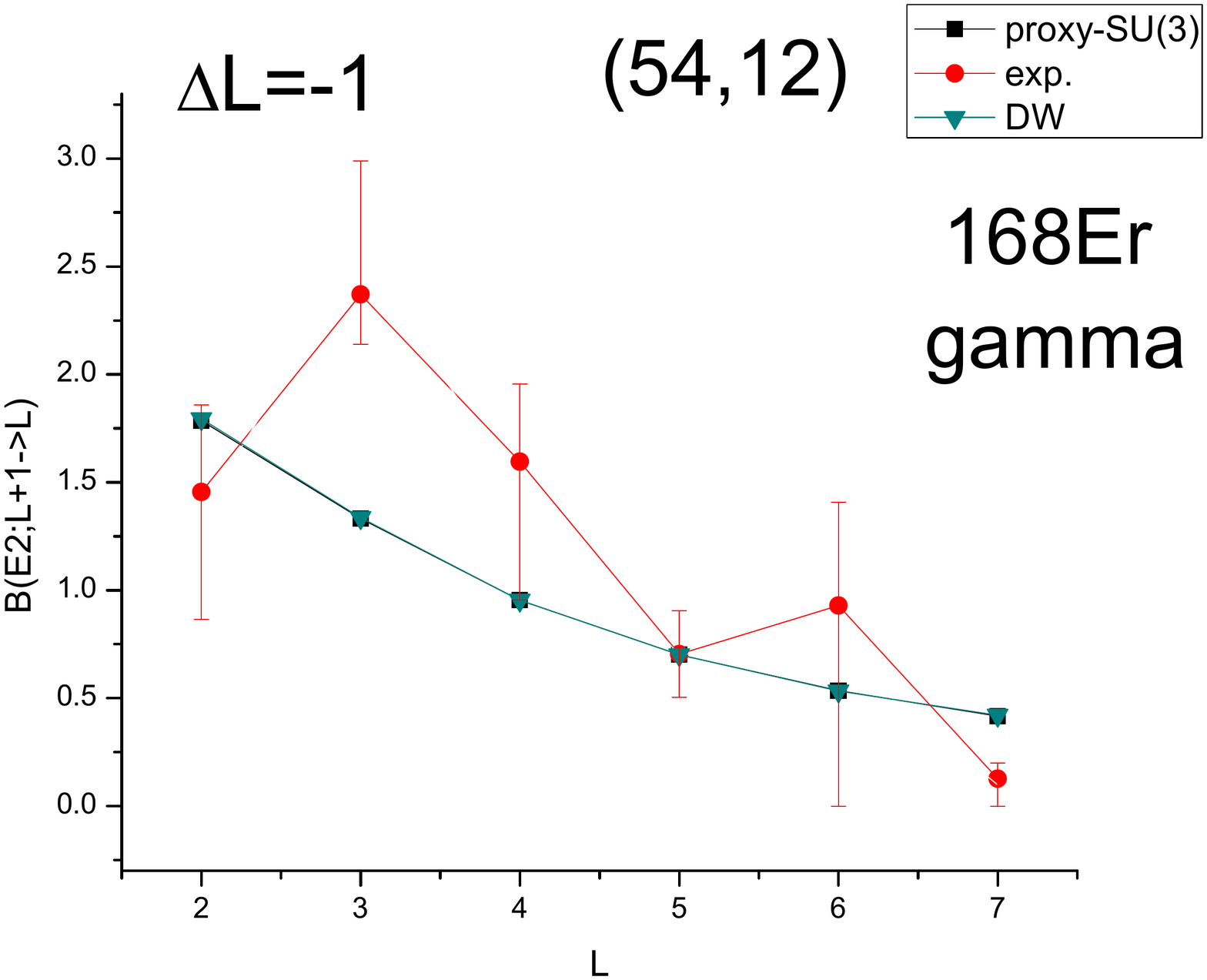}}

\caption{\footnotesize B(E2)s within the ground state band (upper four panels) and within the $\gamma_1$ band (lower four panels) are shown for the indicated proxy-SU(3) irreps and for the relevant nuclei, with data taken from \cite{ENSDF}. In addition, predictions by the pseudo-SU(3) scheme from Refs.
\cite{DW2} (DW), \cite{VV1,VV2,VV3} (VV), and \cite{Kota} (K) are shown. 
All values are normalized to $B(E2; 2_1^+ \to 0_1^+)$. } 
\label{F8}
\end{figure*}

\end{document}